# Creative Quantum Computing: Inverse FFT Sound Synthesis, Adaptive Sequencing and Musical Composition

*Eduardo R. Miranda*

# Creative Quantum Computing: Inverse FFT Sound Synthesis, Adaptive Sequencing and Musical Composition

*Eduardo R. Miranda*

*Interdisciplinary Centre for Computer Music Research (ICCMR)*
*University of Plymouth*
*Ada Lovelace House, 24 Endsleigh Place*
*Plymouth PL4 6DN*
*United Kingdom*

**Abstract:** Quantum computing is emerging as an alternative computing technology, which is built on the principles of subatomic physics. In spite of continuing progress in developing increasingly more sophisticated hardware and software, access to quantum computing still requires specialist expertise that is largely confined to research laboratories. Moreover, the target applications for these developments remain primarily scientific. This chapter introduces research aimed at improving this scenario. Our research is aimed at extending the range of applications of quantum computing towards the arts and creative applications, music being our point of departure. This chapter reports on initial outcomes, whereby quantum information processing controls an inverse Fast Fourier Transform (FFT) sound synthesizer and an adaptive musical sequencer. A composition called *Zeno* is presented to illustrate a practical real-world application.

## 1 Introduction

Quantum computing is emerging as a powerful alternative computing technology, which is built on the principles of subatomic physics. There is a global race for developing quantum computers. A number of them are being developed all over the word.

Fully fledged practical quantum computers are not widely accessible to date. But quantum computing technology is advancing at exponential speeds. Various software simulators are already available [1, 2]. And manufacturers have already started to provide access to quantum hardware via the cloud [3, 4]. These initiatives have enabled experiments with quantum computing to tackle some realistic problems. But much more research is still needed in order to render quantum computers generally useful.

In spite of continuing progress in developing increasingly more sophisticated hardware and software, access to quantum computing still requires specialist expertise that is largely confined to research laboratories. Moreover, the target applications for these developments remain primarily scientific; e.g., quantum chemistry [5] and cryptography [6]. There is a general lack of activity towards widening the range of applications for this alternative computing technology.

The University of Plymouth's Interdisciplinary Centre for Computer Music Research (ICCMR) is championing a new field of research, which we refer to as *Quantum Computer Music*. We are interested in developing quantum computing for musical applications. We are developing bespoke new tools for creating, performing, listening to and distributing music.

This chapter reports an initial practical outcome of our research: a musical composition entitled *Zeno*. *Zeno* is a piece of music for bass clarinet and electronic sounds, whereby a performer interacts with a quantum computer online to create music together live on stage.

We begin with an introduction to the field of Quantum Computer Music followed by a brief survey of related work. Next, we briefly present the topic of algorithmic music systems, which sets the context for the introduction of two quantum music systems of our own design: *qSyn* and *qSeq*. qSyn generates sounds using a quantum hyper-die to control the oscillators of an inverse FFT synthesizer. qSeq is an adaptive melody generator that learns sequencing rules from tunes played to it. Then, we show how these two systems were used to compose *Zeno*. And we demonstrate how quantum computing concepts can inform new technique for musical composition. The chapter ends with concluding remarks.

## 2 Why Quantum Computer Music?

People hardly ever realize that the field of computer music has been progressing in tandem with computer science since the invention of the computer itself. Musicians started experimenting with computing far before the emergence of the vast majority of scientific, industrial and commercial computing applications in existence today.

For instance, as early as the 1940s, researchers at Australia's Council for Scientific and Industrial Research (CSIR) installed a loudspeaker on their Mk1 computer to track the progress of a program using sound. Subsequently, Geoff Hill, a mathematician with a musical background, programmed this machine to playback a tune in 1951 [7].

Then, in the late 1950s composer and Professor of Chemistry, Lejaren Hiller collaborated with mathematician Leonard Isaacson, at University of Illinois at Urbana-Champaign, to program the ILLIAC computer to compose a string quartet entitled *Illiac Suite*. This is often cited as a pioneering piece of algorithmic computer music. That is, whereas Mk1 merely played back an encoded tune, ILLIAC was programmed with algorithms to compose music [8].

Computers play a pivotal part in the music industry today. Emerging alternative computing technology will most certainly have an impact in the way in which we create and distribute music in time to come. Hence the dawn of Quantum Computer Music is a natural progression for music technology.

There have been a number of practice-based musical projects inspired or informed by quantum mechanical processes, but the great majority of these were metaphorical, based on simulation and/or not directly using a quantum computer. Those few projects that did use real-world quantum-related data have been done offline, rather than using physics occurring in real-time during the performance. A notable example is the *LHChamber Music* project [9].

Most probably, the first piece of music that used real-time subatomic physics for performance was *Cloud Chamber* [10]. In *Cloud Chamber*, cosmic rays were made visible and some of them were tracked by vision recognition software and turned in to sound. A violinist played along with this, and in some versions of the performance, the violin sounds triggered a continuous electric voltage that changed the particle tracks, and thus the sounds.

The webpage *Listen to the Quantum Computer Music* is also an interesting initiative [11]. Two tunes are playable online, each of which is a sonification of a well-known quantum algorithm. One is Shor's factorization algorithm [12] and the other is Grover's algorithm for searching a database [13].

By way of related work, we cite initial efforts towards using photonic and adiabatic quantum computers to create music [14]. And James Weaver's pioneering quantum system for generating simple tunes using 17th century music rules [15].

## 3 Algorithmic Music

The first uses of computers in music were for composition. The practice of composition is a tried and tested point of departure for exploration. The great majority of computer music pioneers were composers interested in developing algorithms to generate music. Hence the term 'algorithmic music'. Essentially, the art of algorithmic music consists of harnessing algorithms to produce patterns of data and developing ways to translate them into sound and/or music.

An early approach to algorithmic music, which still remains popular to date, is to program the computer to generate notes randomly and then reject those that do not satisfy given criteria, or rules. Rules based on classic treatises on musical composition are relatively straightforward to encode in a piece of software.

Another widely used approach employs probability distribution functions to predispose the system towards picking specific elements from a given set of musical parameters. For instance, consider the following ordered set of 8 notes that constitute a C3 major scale: {C3, D3, E3, F3, G3, A3, B3, C4} (Figure 1). A Gaussian function would bias the system to pick notes from the middle of the set. That is, it would generate sequences with higher occurrences of F3 and G3 notes.

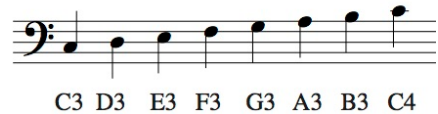

Fig. 1. A given ordered set of musical notes.

A Gaussian function may well be viewed as a simple abstract musical rule. Abstract musical rules can be expressed in a number of ways, including graphs, set algebra, Boolean expressions, finite state automata and transition matrices, to cite but a few. An example using a transition matrix to encode rules for generating sequences of notes is given below. A more detailed introduction to various classic algorithmic music methods is available in [16]. Consider the ordered set shown in Figure 1. Let us define the following rules for establishing which notes are allowed to follow a given note within the set:

    Rule 1: if C3, then either C3, D3, E3, G3 or C4
    Rule 2: if D3, then either C3, E3 or G3
    Rule 3: if E3, then either D3 or F3
    Rule 4: if F3, then either C3, E3 or G3
    Rule 5: if G3, then either C3, F3, G3 or A3
    Rule 6: if A3, then B3
    Rule 7: if B3, then C4
    Rule 8: if C4, then either A3 or B3

Each of these rules represents the transition probabilities for the next note to occur in a sequence. For example, after C3, each of the five notes C3, D3, E3, G3 and C4 has a 20% chance each of occurring. In mathematical terms, each of these notes has the probability $p = 0.2$ to occur. Conversely, notes F3, A3 and B3 will never occur; that is, these notes have the probability $p = 0.0$. In this case the probability has been uniformly distributed between the five candidates, but this does not necessarily need to be uniform.

|      | C3   | D3   | E3   | F3   | G3   | A3   | B3  | C4  |
|------|------|------|------|------|------|------|-----|-----|
| C3   | 0.2  | 0.2  | 0.2  | 0.0  | 0.2  | 0.0  | 0.0 | 0.2 |
| D3   | 0.33 | 0.0  | 0.33 | 0.0  | 0.33 | 0.0  | 0.0 | 0.0 |
| E3   | 0.0  | 0.5  | 0.0  | 0.5  | 0.0  | 0.0  | 0.0 | 0.0 |
| F3   | 0.33 | 0.0  | 0.33 | 0.0  | 0.33 | 0.0  | 0.0 | 0.0 |
| G3   | 0.25 | 0.0  | 0.0  | 0.25 | 0.25 | 0.25 | 0.0 | 0.0 |
| A3   | 0.0  | 0.0  | 0.0  | 0.0  | 0.0  | 0.0  | 1.0 | 0   |
| B3   | 0.0  | 0.0  | 0.0  | 0.0  | 0.0  | 0.0  | 0.0 | 1.0 |
| C4   | 0.0  | 0.0  | 0.0  | 0.0  | 0.0  | 0.5  | 0.5 | 0.0 |

Fig. 2. An example of a transition matrix.

The above rules can be expressed in terms of probability arrays. For instance, the probability array for note C3 is $p$(C3) = [0.2, 0.2, 0.2, 0.0, 0.2, 0.0, 0.0, 0.2] and for note D3 is $p$(D3) = [0.2, 0.0, 0.4, 0.0, 0.4, 0.0, 0.0, 0.0], and so on. The order of the probability coefficients in the array corresponds to the order of the elements in the set. The probability arrays for all the notes of the C3 major scale can be arranged in a two-dimensional matrix, thus forming a transition matrix, as shown in Figure 2.

As computers became increasingly portable and faster, musicians started to program them to create music interactively, during a performance [17]. Let us say, a performer plays a musical note. The computer listens to the note and then produces another one as a response. Most algorithmic music methods that were developed for batch processing of music can be adapted for interactive processing. For instance, given the transition matrix above, if a performer plays the note C4, then the system would respond with A3 or B3.

A sensible approach to get started with research into Quantum Computer Music is to revisit existing algorithmic music methods with a view to running them on quantum computers. Sooner or later new quantum-specific methods are bound to emerge from these experiments; this inspires the methodology that we adopted for our research. Two examples of quantum music systems are introduced below.

## 4 qSyn: Inverse FFT Sound Synthesis

qSyn is an interactive inverse Fast Fourier Transform (FFT) sound synthesizer with parameters supplied by a quantum hyper-die. The system listens to a tune and counts the number of notes. Then, it synthesizes the same amount of sounds as the number of notes that it counted in the tune. The timbres of the synthesized sounds are not homogeneous. And they are not supposed to sound similar to the listened tune in anyway. Their make-up is defined by the quantum hyper-die.

Inverse FFT sound synthesis (also known as additive synthesis) is informed by the theory of Fast Fourier Transform, or FFT. It is based on the notion that the sounds of music can be modelled as a sum of simple sinusoidal waves (Figure 3). Sinusoidal waves are characterized by their respective amplitudes and frequencies. Different values represent perceptible differences in the timbre of the resulting sound. The phases of the sinusoidal waves are also important. For an introduction to sound synthesis methods please refer to [18].

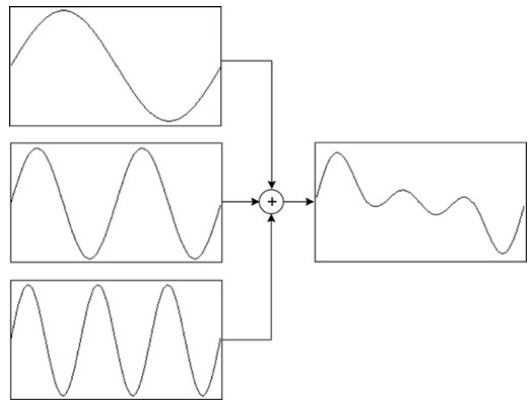

Fig. 3. Example of inverse FFT synthesis of 3 sinusoidal waves.

qSyn comprises 8 oscillators, each of which produces a sinusoidal wave (Figure 4). Each oscillator is controlled by two linear functions, one to handle the amplitude of the produced sinusoid and another to handle its frequency. Phase is not variable in the present version of qSyn; all sinusoids have the same phase.

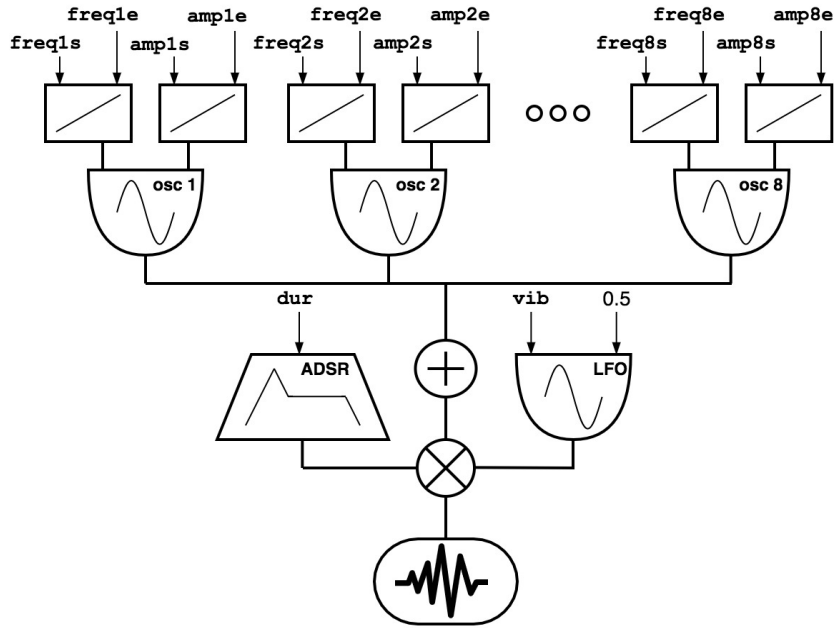

Fig. 4. qSyn's layout.

The linear functions are used to vary amplitudes and frequencies from initial to end values through the duration of the sinusoids. The outputs from the oscillators are summed and a low frequency oscillator (referred to as LFO) is applied to generate a vibrato effect. Vibrato blends the sinusoids, rendering the result more realistic to our ears. Then an ADSR function, or envelope, shapes the overall amplitude of the sound. The input parameters for qSyn are listed in Table 1.

At the core of the quantum hyper-die is a simple quantum circuit that uses the Hadamard gate to puts 9 qubits in superposition and measures them (Figure 5). (An introduction to the fundamentals of quantum computing is beyond the scope of this chapter. Please consult [19].) These result in a set of 9 measurements, which is fed into an algorithm that calculates binary triplets. Then, these triplets are used to retrieve parameters for the synthesizer.

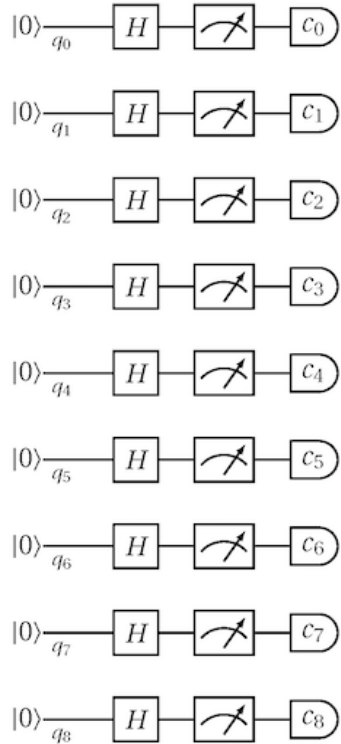

Fig. 5. The quantum hyper-die circuit. The operator, or gate H (for Hadamard) puts all 9 qubits in superposition.

Consider the list of measurements $C = [c_8, c_7, c_6, c_5, c_4, c_3, c_2, c_1, c_0]$ and a given list of 8 frequencies $Freq = [f_0, f_1, f_2, f_3, f_4, f_5, f_6, f_7]$. The triplets are formed by combining 3 elements from the list $C$. For example, $(c_8\ c_7\ c_6)$, $(c_6\ c_7\ c_8)$, $(c_5\ c_4\ c_3)$, $(c_3\ c_4\ c_5)$, $(c_2\ c_1\ c_0)$, $(c_0\ c_1\ c_2)$ and so forth. The decimal value of a triplet gives an index to retrieve a frequency $f_n$ from the list *Freq*. For instance, the triplet (0 1 0), which gives the decimal number 2, would retrieve $f_2$.

Each qSyn's synthesis parameter is coupled with a unique triplet formation. For instance, ($c_8$ $c_7$ $c_6$) is coupled with the starting frequency for oscillator number 1 and ($c_6$ $c_7$ $c_8$) with the ending frequency for this oscillator. And ($c_5$ $c_4$ $c_3$) is coupled with the starting frequency for oscillator number 2, and so on (see Table 1. Durations, not shown on this table, are specified in seconds.). In other words, the decimal values of the triplets are used to retrieve respective parameters for the synthesizer. The system holds a database of frequencies, amplitudes and durations, which can be customized.

For each sound to be produced, the hyper-die circuit is invoked twice: once to produce triplets for frequencies and then again to produce triplets for amplitudes and the duration of the sound.

As an example, let us consider that the system is prompted to synthesize 4 sounds. And let us assume that there are 8 arrays of frequencies ($Freq^n$) available in the database, one for each oscillator. And there is an array of amplitudes *Amp*, which serves all oscillators, as follows:

$Freq^1$ = [55.0, 277.18, 220.0, 329.63, 164.81, 277.18, 220.0, 329.63]
$Freq^2$ = [82.4, 369.99, 293.67, 196.0, 466.16, 369.99, 293.67, 196.0]
$Freq^3$ = [87.3, 349.23, 277.18, 440.0, 87.3, 349.23, 277.18, 440.0]
$Freq^4$ = [92.49, 415.3, 329.63, 233.08, 523.25, 329.63, 233.08, 523.25]
$Freq^5$ = [435.53, 1468.32, 1038.26, 1959.97, 2330.81, 1468.32, 1038.26, 1959.97]
$Freq^6$ = [440.0, 2217.46, 1760.0, 2637.02, 1318.51, 2217.46, 1760.0, 2637.02]
$Freq^7$ = [435.53, 1746.14, 1385.91, 2200.0, 435.53, 1746.14, 1385.91, 2200.0]
$Freq^8$ = [741.66, 2354.63, 1571.52, 3143.05, 3960.0, 2354.63, 1571.52, 3143.05]

*Amp* = [0.06, 0.08, 0.1, 0.12, 0.14, 0.16, 0.18, 0.2]

In order to synthesize, let us say, 4 sounds, the system would need to run the quantum hyper-die circuit 8 times: 4 times to produce a set of measurements $\mathcal{C}$ to retrieve frequencies and 4 times to produce a set of measurements $\mathcal{D}$ to retrieve amplitudes and durations. Possible results are as follows:

$$\mathcal{C} = \{[0, 0, 0, 0, 0, 1, 0, 0, 1], [0, 1, 1, 1, 1, 1, 1, 0, 1, 0],$$
$$[0, 0, 1, 0, 1, 1, 1, 1, 1], [1, 1, 1, 0, 1, 0, 0, 1, 1]\}$$

$$\mathcal{D} = \{[0, 0, 1, 0, 1, 1, 0, 0, 0], [1, 0, 1, 0, 1, 1, 1, 1, 0],$$
$$[1, 1, 0, 1, 1, 1, 0, 0, 0], [0, 0, 1, 1, 1, 0, 1, 0, 0]\}$$

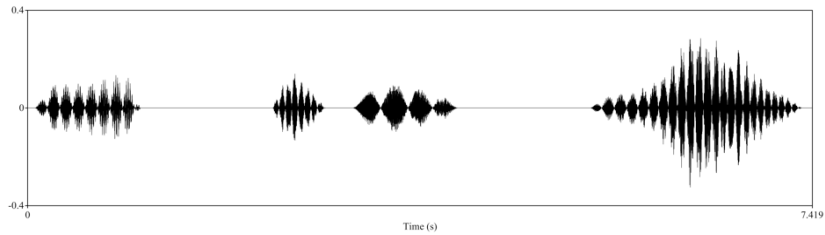

Fig. 6. Four sounds produced by qSyn.

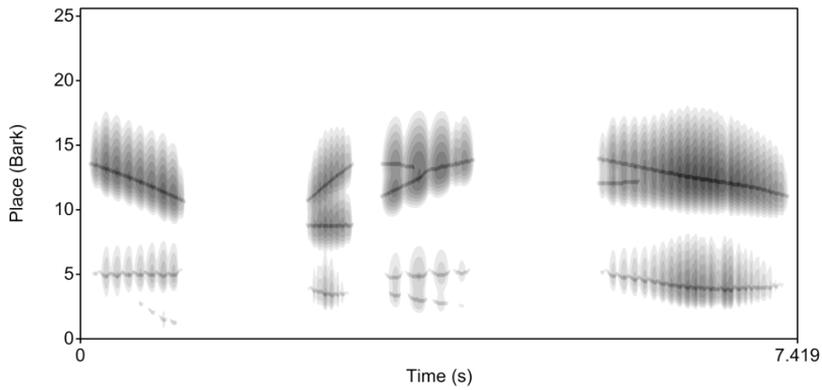

Fig. 7. The cochleagram reveals salient spectral components sliding upwards or downwards.

Next, the system calculates the triplets and retrieves the parameters for the synthesizer. For instance, in the first run the triplet ($x_0$ $x_1$ $x_2$) is equal to 000. Therefore, it retrieves the first element of *Freq*[1] for the starting frequency of oscillator 1, which is 55.0 Hz. Table 1 shows the retrieved parameters for the first sound.

## 5 qSeq: Adaptive Musical Sequencer

qSeq is an adaptive musical sequencer that listens to a tune and responds with a sequence of notes. The response is created based on an analysis of the tune.

The system extracts 3 features from the tune: the pitches of the notes, their durations and their loudness. With this information, the system builds transition matrices representing how often those features occurred in the tune.

It could be said at this point that qSeq learns the style of the tune. As we have seen earlier, transition matrices embody rules for producing sequences.

Next, the matrices are converted into a list of angles [$\vartheta 1, \vartheta 2, ... \vartheta 18$], which are used to rotate qubits in a quantum circuit representing transition probabilities in terms of quantum states (Figure 8). This is a further development of a circuit proposed by James Weaver [15] to generate a musical sequence. In Figure 8, the quantum gate RY($\vartheta$) rotates a qubit by a given angle $\vartheta$ around the y-axis on the Bloch sphere.

The circuit is armed and measured as many times as the required number of notes to be produced. Each time the circuit is measured it generates a note. For instance, if qSeq is to generate 20 notes, the circuit is armed and measured 20 times. To start with, all qubits are put in superposition (with the gate H) before being processed by the rest of the circuit. Then, after the first round, the resulting measurements are used to arm the qubits for the circuit, as shown schematically in Figure 9. For clarity, the dashed portion of the circuit in Figure 9 is encapsulated into a mega quantum gate labelled as 'Transition'. Figure 9 shows the measurements being fed back into the circuit to compute the next note.

Table 1. Parameters for synthesizing a sound. Frequency of vibrato, duration of the sound and the silence interval between the sounds are now shown. Frequencies are given in Hertz. qSyn works with values between 50 Hz and 8 kHz. Amplitudes are specified as a number between 0.0 (no signal) and 1.0 (maximum power).

| Triplet Code | Binary | Decimal | Parameter | Retrieved Value |
|---|---|---|---|---|
| $(c_8\ c_7\ c_6)$ | 000 | 0 | `freq1s` | 55.0 Hz |
| $(c_6\ c_7\ c_8)$ | 000 | 0 | `freq1e` | 55.0 Hz |
| $(c_5\ c_4\ c_3)$ | 001 | 1 | `freq2s` | 369.99 Hz |
| $(c_3\ c_4\ c_5)$ | 100 | 4 | `freq2e` | 466.16 Hz |
| $(c_2\ c_1\ c_0)$ | 001 | 1 | `freq3s` | 349.23 Hz |
| $(c_0\ c_1\ c_2)$ | 100 | 4 | `freq3e` | 87.3 Hz |
| $(c_7\ c_6\ c_5)$ | 000 | 0 | `freq4s` | 92.49 Hz |
| $(c_5\ c_6\ c_7)$ | 000 | 0 | `freq4e` | 92.49 Hz |
| $(c_4\ c_3\ c_2)$ | 000 | 0 | `freq5s` | 435.53 Hz |
| $(c_2\ c_3\ c_4)$ | 000 | 0 | `freq5e` | 435.53 Hz |
| $(c_8\ c_5\ c_2)$ | 000 | 0 | `freq6s` | 440.0 Hz |
| $(c_2\ c_5\ c_8)$ | 000 | 0 | `freq6e` | 440.0 Hz |
| $(c_7\ c_4\ c_3)$ | 011 | 3 | `freq7s` | 2200.0 Hz |
| $(c_3\ c_4\ c_7)$ | 110 | 6 | `freq7e` | 1385.91 Hz |
| $(c_6\ c_3\ c_0)$ | 001 | 1 | `freq8s` | 2354.63 Hz |
| $(c_0\ c_3\ c_6)$ | 100 | 4 | `freq8e` | 3960.0 Hz |
| $(d_8\ d_7\ d_6)$ | 001 | 1 | `amp1s` | 0.08 |
| $(d_6\ d_7\ d_8)$ | 100 | 4 | `amp1e` | 0.14 |
| $(d_5\ d_4\ d_3)$ | 011 | 3 | `amp2s` | 0.12 |
| $(d_3\ d_4\ d_5)$ | 110 | 6 | `amp2e` | 0.18 |
| $(d_2\ d_1\ d_0)$ | 000 | 0 | `amp3s` | 0.06 |
| $(d_0\ d_1\ d_2)$ | 000 | 0 | `amp3e` | 0.06 |
| $(d_7\ d_6\ d_5)$ | 010 | 2 | `amp4s` | 0.1 |
| $(d_5\ d_6\ d_7)$ | 010 | 2 | `amp4e` | 0.1 |
| $(d_4\ d_3\ d_2)$ | 000 | 0 | `amp5s` | 0.06 |
| $(d_2\ d_3\ d_4)$ | 000 | 0 | `amp5e` | 0.06 |
| $(d_8\ d_5\ d_2)$ | 010 | 2 | `amp6s` | 0.1 |
| $(d_2\ d_5\ d_8)$ | 010 | 2 | `amp6e` | 0.1 |
| $(d_7\ d_4\ d_3)$ | 110 | 6 | `amp7s` | 0.18 |

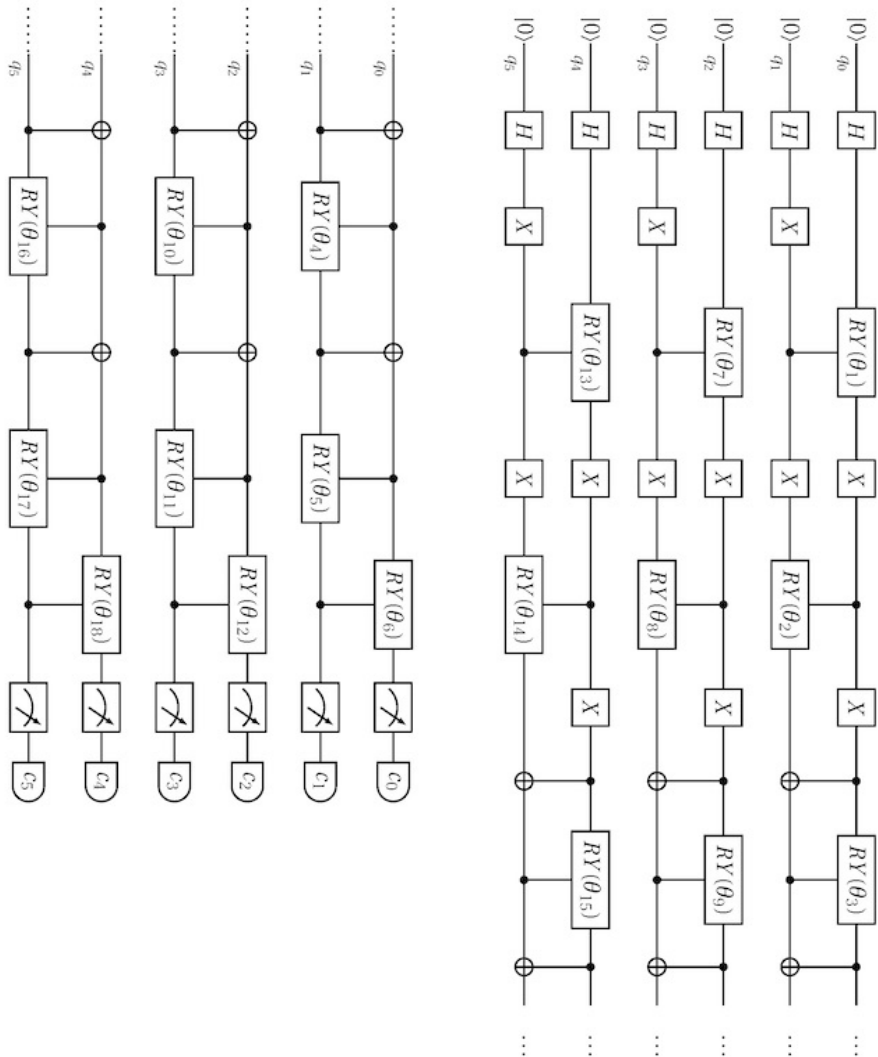

Fig. 8. qSeq quantum circuit.

Figure 6 plots the resulting sound and Figure 7 shows its respective cochleagram. Notice in the cochleagram the effect of the synthesizer's linear functions on the spectrum of the sounds. Salient components of the spectrum are clearly shown sliding upwards or downwards.

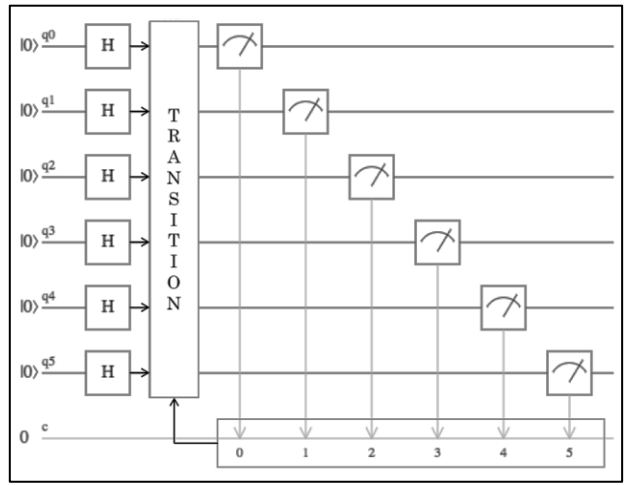

Fig. 9. qSeq with encapsulated Transition mega gate.

As a demonstration, let us consider the case where the system listened to the tune shown in Figure 10, which is the opening of Beethoven's 5th symphony.

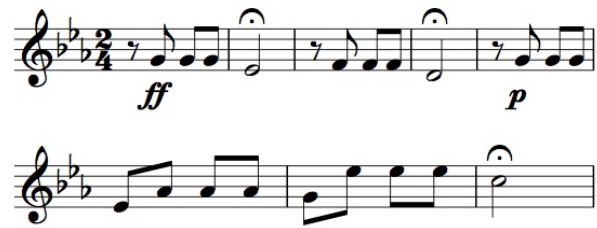

Fig. 10. The opening of Beethoven's 5th symphony.

Even though the quantum circuit in Figure 8 can represent the style of the original tune, the notes generated by the system do not necessarily resemble the ones that were used to build the matrices. To put it in very simplistic terms, imagine a situation where at the learning stage a system learns rules, such as the ones shown in Figure 2. But then, at the generative stage the system substitutes the notes by a different set of notes, preserving the probability coefficients. In music jargon, we say that the system learned *musical form* rather than musical content.

In this case the system extracted the following information ($P$ = pitches, $D$ = durations and $L$ = loudnesses):

$P$ = [67, 67, 67, 63, 65, 65, 65, 62, 67, 67, 67, 63, 68, 68, 68, 67, 75, 75, 75, 72]

$D$ = [298, 301, 302, 1798, 302, 297, 301, 1799, 302, 303, 296, 302, 297, 302, 298, 301, 297, 301, 297, 1799]

$L$ = [113, 113, 113, 105, 113, 113, 113, 107, 61, 61, 61, 57, 64, 63, 63, 61, 70, 68, 67, 60]

Pitches and loudnesses are represented in the terms of MIDI codes [21] and durations in terms of milliseconds. The system uses 4 x 4 matrices; this is a limitation imposed by the qSeq circuit. Higher dimensions would require circuits with greater depth, which increases the effect of quantum decoherence [19]. Decoherence must be avoided because it causes computation errors. Improved quantum hardware and better error correction methods will enable circuits with greater depth in the future.

Therefore, the amount of musical information needs to be reduced here. In order to do this the system processes only the first four different elements of each list. For instance, the first four different elements in $P$ are 67, 63, 65 and 62. Thus, it keeps these in the list and removes all others. The results are as follows:

$P'$ = [67, 67, 67, 63, 65, 65, 65, 62, 67, 67, 67, 63, 67]

$D'$ = [298, 301, 302, 1798, 302, 301, 302, 302, 302, 298, 301, 301]

$L'$ = [113, 113, 113, 105, 113, 113, 113, 107, 61, 61, 61, 61]

Next, the system builds matrices counting the number of times a certain element in the horizontal axis followed another in the vertical axis. For instance, pitch 63 followed pitch 67 twice, whereas pitch 65 followed 63 only once (Figure 11).

| P' | 67 | 63 | 65 | 62 |
|----|----|----|----|----|
| 67 | 4 | 2 | 0 | 0 |
| 63 | 1 | 0 | 1 | 0 |
| 65 | 0 | 0 | 2 | 1 |
| 62 | 1 | 0 | 0 | 0 |

Fig. 11. Matrix for pitch, counting the number of times a certain element in the horizontal axis followed an element in the vertical axis.

The next step is to convert each of the 3 counting matrices into bistochastic matrices. Figure 12 shows the bistochastic matrix for pitches.

| P' | 67 | 63 | 65 | 62 |
|----|----|----|----|----|
| 67 | 0.131 | 0.858 | 0.001 | 0.009 |
| 63 | 0.192 | 0.03 | 0.724 | 0.053 |
| 65 | 0.001 | 0.005 | 0.244 | 0.750 |
| 62 | 0.676 | 0.106 | 0.031 | 0.188 |

Fig. 12. Bistochastic matrix for note pitches.

Then, 3 identity matrices are created, defining vector spaces with 6 degrees of freedom each. Bistochastic and respective identity matrices are gradually rotated until the difference between the corresponding entries of each matrix are minimized. The aim is to perform these rotations on 6-dimensional vector spaces, thus yielding 6*3=18 angles for the circuit.

The resulting angles for each degree-of-freedom rotation are the angles $\vartheta$ for the RY gates of the quantum circuit (Figure 8). The resulting angles in degrees for our example are as follows:

$\Theta P' = [243, 197, 243, 186, 180, 249]$
$\Theta D' = [237, 203, 128, 203, 169, 249]$
$\Theta L' = [220, 180, 157, 180, 140, 123]$

The complete list of angles $[\vartheta 1, \vartheta 2, \ldots \vartheta 18]$ for the 18 RY gates of the quantum circuit shown in Figure 8 is achieved by concatenating $\Theta P' \sqcup \Theta L' \sqcup \Theta O'$. Thus, $\Theta = [243, 197, 243, 186, 180, 249, 237, 203, 128, 203, 169, 249, 220, 180, 157, 180, 140, 123]$.

For the generative sequencing process, to begin with, the states of the 6 qubits are put in superposition. Then, at subsequent rounds, the circuit is armed with qubits in states echoing the measurements produced in the most recent round. Thus, if a given round produced $C = [c_5, c_4, c_3, c_2, c_1, c_0]$, then the circuit will be armed for the next round with $|c_4\rangle|c_5\rangle|c_2\rangle|c_3\rangle|c_0\rangle|c_1\rangle$. The reason for inverting consecutive pairs of measurements for arming the qubits (e.g., $|c_4\rangle|c_5\rangle\ldots$" instead of $|c_5\rangle|c_4\rangle\ldots$") is due to the way which quantum circuits are normally implemented: the results in the list of measurements $C$ are ordered back to front.

In practice, the qSeq circuit consists of 3 independent circuits operating concurrently, each of which produces a pair of measurements. The initial states of each concurrent circuit's pair of qubits represent input parameters to generate the next note in a sequence. Hence the respective pairs are inverted to denote the codes of those musical parameters (see Figures 13 and 14).

For instance, consider the first 4 rounds of our example:

Round 1:
Initial default quantum state: H($|0\rangle$) $\otimes$ H($||0\rangle$) $\otimes$ H($||0\rangle$) $\otimes$ H($|0\rangle$) $\otimes$ H($|0\rangle$) $\otimes$ H$|0\rangle$)
Results from measurements: $C = [0, 1, 1, 0, 1, 0]$

Round 2:
Quantum state: $|1\rangle|0\rangle|0\rangle|1\rangle|0\rangle|1\rangle$
Results from measurements: $C = [1, 1, 1, 1, 0, 0]$

<u>Round 3</u>:
Quantum state: $|1\rangle|1\rangle|1\rangle|1\rangle|0\rangle|0\rangle$
Results from measurements: $C = [0, 0, 0, 0, 0, 0]$

<u>Round 4</u>:
Quantum state: $|0\rangle|0\rangle|0\rangle|0\rangle|0\rangle|0\rangle$
Results from measurements: $C = [1, 0, 1, 0, 1, 0]$

The measurements from each round embody binary codes that are used to generate a musical response. In order to generate a note, the system retrieves its pitch from a given set of pitches (Figure 14) and its duration from a given set of durations (Figure 13).

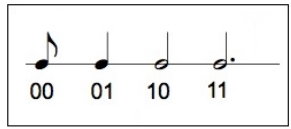

Fig. 13. The set of durations contains an eight note (1/2 beat), a quarter note (1 beat), a half note (2 beats) and a dotted half note (3 beats).

The system holds a set of 4 durations and a set of 16 pitches grouped into 4 subsets of 4 pitches each. The durations, subsets, and pitches are tagged with binary codes, which are used to retrieve them.

Let us consider the result from first round of measurements shown above: $C = [0, 1, 1, 0, 1, 0]$. The binary codes are defined as follows:

$(c_0\ c_1)$ = code to establish the set from which the pitch of a note will be
　　　　retrieved
$(c_4\ c_5)$ = code to retrieve the pitch of a note
$(c_2\ c_3)$ = code to retrieve the duration of note

In this case, code $(c_0\ c_1) = 01$ indicates that pitch will be retrieved from set B, and code $(c_4\ c_5) = 10$ determines that the pitch of this note is E$\flat$. And the code $(c_2\ c_3) = 01$ determines that this is a quarter note. Next, the measurements $C = [1, 1, 1, 1, 0, 0]$ produce a dotted half G note from pitch set A. And so on. Let us assume that the system was prompted to produce 20 notes in response to the Beethoven tune in Figure 10. The result is shown in Figure 15. In this case, code $(c_0 \ c_1) = 01$ indicates that pitch will be retrieved from set B, and code $(c_4 \ c_5) = 10$ determines that the pitch of this note is E♭. And the code $(c_2 \ c_3) = 01$ determines that this is a quarter note. Next, the measurements $C = [1, 1, 1, 1, 0, 0]$ produce a dotted half G note from pitch set A. And so on. Let us assume that the system was prompted to produce 20 notes in response to the Beethoven tune in Figure 10. The result is shown in Figure 15.

Fig. 14. Set of pitches. Although the pitches in the set are notated as half notes, or minims, the actual duration of a note in the response is picked from the set of durations shown in Figure 13.

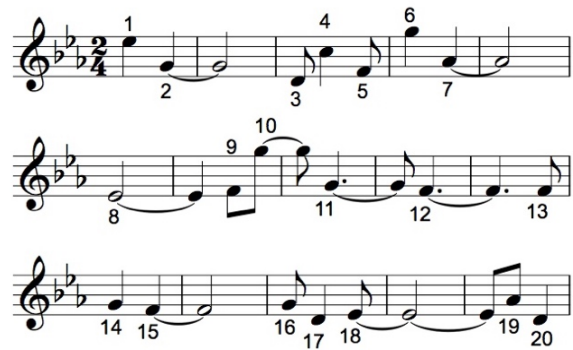

Fig. 15. A musical response produced by qSeq. The durations of two or more notes of identical pitches linked by an arc are summed. They are considered as if they were a single note. This normally happens in music notation when one needs to write a note that will last longer than the boundary imposed by the time signature. For instance, note number 2 is a single sound, whose duration crosses over the bar, whereas notes 12 and 13 are two separate ones.

## 6 The Composition *Zeno*

*Zeno* is a musical composition of for bass clarinet and electronic sounds. The first two pages of the musical score representing how the piece is structured is shown in the Appendix. The electronics comprise (a) recordings of pre-composed sounds (i.e., the 'samples' mentioned on the score), which are played back during the performance, and (b) sounds generated interactively with a quantum computer. The latter is the focus of interest here: during the performance, the system listens to the bass clarinet and generates responses accordingly.

The interactive system consists of two components: a client and a server. The client runs on a standard laptop computer and the server runs on Rigetti's Forest quantum computer, which is located in Berkeley, California. The server runs the two quantum circuits introduced above. And the client takes care of the music and sound processing tasks.

The client listens to musical phrases via a microphone. Then, it extracts information from the audio signal, prepares the data and relays them to the server. Together with the data relayed to the server, the client also sends an instruction indicating which circuit to run. The server receives the data, runs the required quantum circuit, and relays the measurement results back to the client. Next, the client processes the measurements to synthesize sounds or playback note sequences, depending on which circuit the server ran. Upon receiving the measurements, the client then activates the respective generative algorithm – and, if required, a synthesizer - to produce the response (Figure 16).

The musical score contains performance instructions for the performer and pre-defines which circuit to use and when. In addition to prescribing pre-set musical phrases to be played on the bass clarinet, there are moments where the performer is asked to improvise with a given set of musical notes. It is these improvisations that are listened to and processed by the system. For instance, the extract in Figure 17 shows as instance where the performer is asked to improvise with a given set of 7 notes for 5 seconds. Here the system is programmed to enter in listening mode for 5 seconds (indicate by a box labelled as '#5 QC Listening' on the lower stave). The listened sound is processed, the client and server perform their jobs, and a sound response is produced. In this case, the sounds are produced by qSyn and the duration of the response is not known in advance; this depends on the numbers of notes played by the performer and the measurements of the quantum hyper-die. As soon as the response starts, the performer is asked to pick notes from set A or B and improvise with them along with qSyn's response.

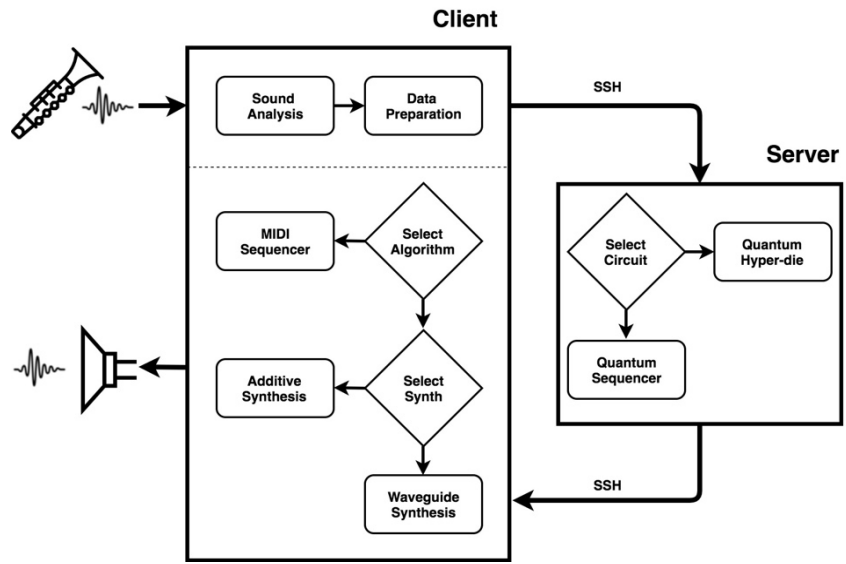

Fig. 16. The system's layout for *Zeno*. In addition to the qSyn introduced earlier, the client also runs a synthesizer that simulates the sounds of the clarinet using the Waveguides synthesis method [18]; it also handles codes calculated from the hyper-die's measurements to retrieve synthesis values.

There is a short waiting period between the moment when the listening mode runs out of time and the musical response. This is because the system needs time to process audio and music data. And it takes time to handle data through the Internet. In order to avoid breaking the flow of the music, the performer may continue the improvisation after the listening mode has ran out. But the system will not process anything that is played when it is not listening.

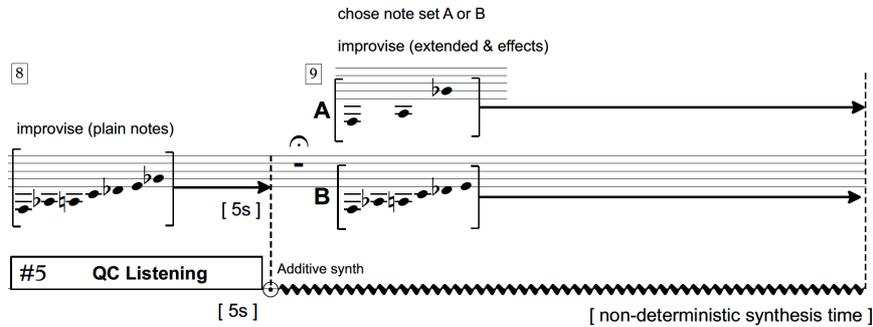

Fig. 17. Extract from the musical score showing an example of performer's choice.

Monophonic musical instruments, such as the flute or the clarinet, produce 1 note at a time. However, skilled players are able to produce multiphonics on these instruments; that is, more than one note at once. On a wind musical instrument, this is can be achieved by blowing air through the device in such a way that the resulting spectrum is split around more than one prominent, or fundamental, frequency. The bass clarinet is an interesting case in point [20].

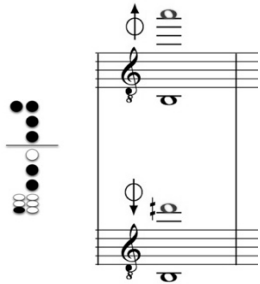

Fig. 18. Two distinct multiphonics variants are produced with the same fingering position on the keys of the bass clarinet. The fingering position is shown on the left side of the figure; black circles indicate the keys of the instrument that need to be pressed.

It is possible to play two distinct multiphonics variants with the same fingering position on the keys of the bass clarinet. But it is often the case that performers cannot predict accurately which of the two sets of multiphonics will result. Figure 18 shows an example of this. Given a base note B2, the same fingering position produces either an additional A5 (top stave) or a slightly sharped D4 (bottom stave).

Whereas the unstable nature of such multiphonics might be regarded as an inconvenience, they actually come in handy for a 'quantum-compliant' musical scenario.

Let us imagine a situation where a bass clarinet performer encounters an instruction to produce multiphonics with a specific fingering position. In a way, the act of following this instruction is analogous to measuring a quantum system: the performer is not certain which multiphonics will be produced until she actually plays the instrument. Even if the performer steers her efforts towards producing certain multiphonics, this would only increase the probability of obtaining them. Metaphorically, the multiphonics for that specific fingering position are in superposition until a performer plays them; or 'observe' them, in quantum mechanics parlance. The extract in Figure 19 shows an example of how this phenomenon is used in *Zeno* to create a quantum-compliant compositional technique.

In Figure 19, the performer is asked to play multiphonics using a specific fingering position. The possible two top multiphonics are shown within brackets on the first chord at left side of the figure. After a brief moment of uncertainty (indicated by the wavy live), the multiphonics settle into one or the other, indicated as "a" or "b". After a few moments, the performer than picks either note set A or B to improvise with. However, in this case, this choice is dictated by the multiphonics. If the multiphonics settled with the higher note A5 then the performer improvises for 15 seconds with notes from set A. Otherwise she improvises with notes from set B. The system listens to this improvisation and then synthesizes a sound. Here qSyn does so using the Waveguides synthesis method (Figure 16).

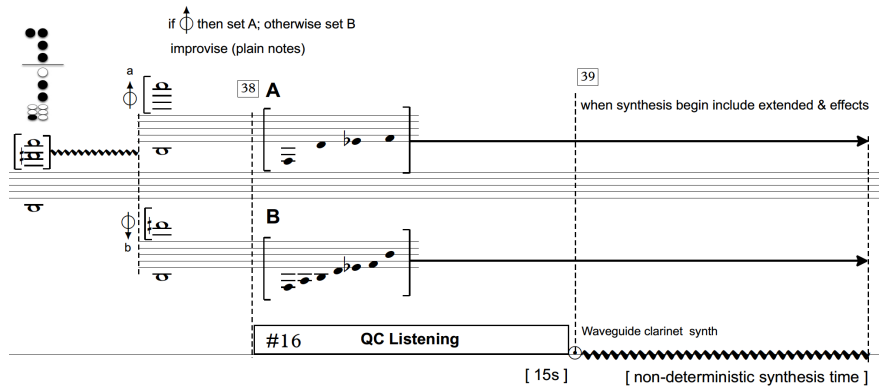

Fig. 19. Extract from the musical score showing an example using multiphonics.

An example of a section using qSeq is shown in Figure 20. In this case, the system listens to the performer improvising with notes from a given set for 30 seconds. Then, it processes the listened sequence, as explained in the qSeq section above, and generates a response lasting for 1 minute. The system's response is a sequence of musical notes encoded as MIDI information [21]. MIDI is a protocol that allows computers, musical instruments and other hardware to communicate.

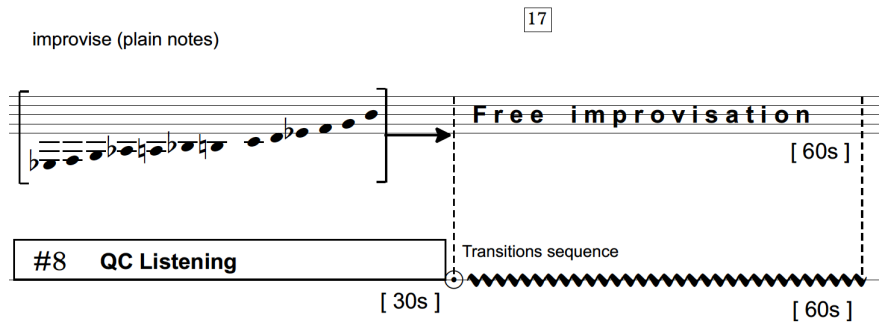

Fig. 20. Extract from the musical score with free improvisation alongside materials generated by qSeq.

The difference of encoding the results with MIDI rather than synthesizing sounds is that we can plug in third-party music software to visualize or play back the sequence. For instance, the score in Figure 15 was produced simply by uploading the resulting MIDI data into a music score editor.

For *Zeno*, MIDI data are channeled straight into a stack of electronic music instruments, drum machines and samplers; a sampler is a digital musical instrument whose sounds are recordings stored in its memory. Each MIDI note triggers a few of these in polyphony.

## 7 Concluding Remarks

In spite of continuing progress in developing increasingly more sophisticated hardware and software for quantum computing, access to this technology still requires specialist expertise that is largely confined to research laboratories. Moreover, the target applications for these developments remain primarily scientific; e.g., quantum chemistry and cryptography.

Yet, computers are essential for almost every domain of human endeavor these days. It is therefore surprising that there have been virtually no initiatives to date towards widening the range of uses for quantum computers beyond scientific applications. In particular applications for the entertainment industry and creative economies.

In order to address the lack of activity beyond scientific applications of quantum computing, the University of Plymouth's Interdisciplinary Centre for Computer Music Research is developing a new field of research, which we refer to as *Quantum Computer Music*. We are developing musical applications of quantum computing, which include bespoke new tools and approaches to creating, performing, listening to and distributing music.

This article introduced as initial outcome from our research, which comprises the implementation of software and a practical musical composition whereby a performer interacts with a quantum computer to create music during a performance.

The mainframe used to compose the pioneering *Illiac Suit*e was one of the first computers built in the USA, comprising thousands of vacuum tubes. These early computers generated thousands of watts of heat. They needed to be housed in rooms with heavy-duty air conditioning systems. This sounds somewhat familiar to quantum computer engineers, in the sense that current quantum processors have to be kept to near absolute zero temperatures, under minus 270 Celsius, to operate.

It is often said that today's quantum computers are in a development stage comparable to those clunky mainframes built in the mid of the last century. By the same token, as early pioneers of computer music paved the way for the development of a thriving global music industry, today we are on a mission to develop quantum computing technology for music.

Admittedly, the two quantum systems introduced above could as well be implemented on standard digital computers. At this stage, we are not advocating any quantum advantage for musical applications. What we advocate, however, is that the music technology community should be quantum-ready for when quantum computing technology becomes more sophisticated, widely available, and possibly advantageous for creative tasks. In the process of learning and experimenting with this new technology, novel approaches and innovative creative ideas will certainly emerge. A glimpse of this is the quantum-compliant composition technique with multiphonics that we invented for *Zeno*.

## A.1. Appendix

The first two pages of *Zeno*'s musical score. The electronics part is represented as a squiggly line under the bass clarinet part; for instance, sample 01 lasts for 75 seconds, sample 02 for 6 seconds, and so on.

Zeno

Eduardo Reck Miranda
Sep 2019

## Acknowledgements

The author would like to thank Rigetti Computing, in California, for supporting our research with priority access to quantum computing hardware and technical advice. Special thanks to Rigetti's Amy Brown, Tushar Mittal and Tom Lubowe for fruitful discussions. Many thanks to bass clarinetist Sarah Watts for sharing her knowledge on multiphonics and contributing to the composition of *Zeno*. Also, thanks to James Hefford (University of Oxford), James R. Wootton (IBM) and James Weaver (IBM) for comments and suggestions.